\documentclass[12 pt]{article} 
\usepackage{times,bm}
\usepackage[affil-it,]{authblk}
\usepackage[pagebackref=false,colorlinks=true,linkcolor=blue,citecolor=blue,urlcolor=blue]{hyperref}
\usepackage{geometry}
\usepackage{graphicx}
\usepackage{cite}
\usepackage{amsmath}
\usepackage{amsfonts}
\usepackage{amssymb}
\usepackage{color}
\usepackage{float}
\usepackage{multirow,multicol}
\usepackage{tabulary}
\usepackage[flushleft]{threeparttable}
\usepackage{pgfplots,subfigure}
\usepackage{xcolor}
\numberwithin{equation}{section}
\usepackage{tikz}
\usepackage{ulem}
\usepackage{hyperref}
\usepackage{nameref}
\usepackage{comment}

\usepgflibrary{arrows}
\usetikzlibrary{shapes.callouts}
\tikzset{  
	level/.style   = { thick, },
	connect/.style = { dotted, red   },
	notice/.style  = { draw, rectangle callout, callout relative pointer={#1} },
	label/.style   = { text width=2cm }
}

\usepackage{letltxmacro}
\makeatletter
\let\oldr@@t\r@@t
\def\r@@t#1#2{%
	\setbox0=\hbox{$\oldr@@t#1{#2\,}$}\dimen0=\ht0
	\advance\dimen0-0.2\ht0
	\setbox2=\hbox{\vrule height\ht0 depth -\dimen0}%
	{\box0\lower0.4pt\box2}}
\LetLtxMacro{\oldsqrt}{\sqrt}
\renewcommand*{\sqrt}[2][\ ]{\oldsqrt[#1]{#2}}
\makeatother
\pgfplotsset{compat=1.17}
\begin{document}

\newcommand{{\ri}}{{\rm{i}}}
\newcommand{{\Psibar}}{{\bar{\Psi}}}

\title{Thermodynamics of the Schwarzschild and Reissner–Nordstr\"{o}m black holes under higher-order generalized uncertainty principle}
\author{\large \textit {S. Hassanabadi}$^{\ 1}$\footnote{E-mail: s.hassanabadi@yahoo.com}~,~ \textit {J. K\v{r}\'i\v{z}}$^{\ 1}$\footnote{E-mail: jan.kriz@uhk.cz}~,~\textit {W. S. Chung}$^{\ 2}$\footnote{E-mail: mimip44@naver.com}~,~\textit {B. C. L\"{u}tf\"{u}o\u{g}lu}$^{\ 1\,, 3}$\footnote{E-mail: bclutfuoglu@akdeniz.edu.tr (corresponding author)}~,~\textit {E. Maghsoodi}$^{\ 4}$\footnote{E-mail: e.maghsoodi184@gmail.com} ~and~\textit {H. Hassanabadi}$^{\ 1\,,5}$\footnote{E-mail: h.hasanabadi@shahroodut.ac.ir }  \\

	\small \textit {$^{\ 1}$Department of Physics, University of Hradec Kr\'alov\'e,Rokitansk\'eho 62, 500 03 Hradec Kr\'alov\'e, Czechia}\\
	\small \textit {$^{\ 2}$Department of Physics and Research Institute of Natural Science, College of Natural Science, Gyeongsang National University, Jinju, Korea}\\
	\small \textit {$^{\ 3}$Department of Physics, Akdeniz University, Campus 07058, Antalya, Turkey}\\
	\small \textit {$^{\ 4}$Department of Physics, Faculty of Science, Lorestan University, Khorramabad, Iran}\\
	\small \textit {$^{\ 5}$Faculty of Physics, Shahrood University of Technology, Shahrood, Iran}\\
	\small \textit{P.O. Box 3619995161-316}\\
}

\date{}
\maketitle

\begin{abstract}		
{In this manuscript, the modification of thermodynamic properties of Schwarzschild and Reissner-Nordstr\"{o}m black holes in the presence of higher-order generalized uncertainty principle (GUP) have been investigated. We considered heuristic analysis versus the behavior of a particle that is absorbed by the black hole, and then, we studied the thermodynamic properties of the black holes such as temperature-mass, temperature-charge-mass, entropy, and heat capacity functions up to the first and second-order of the GUP parameter. We compared the behaviors the GUP-corrected thermodynamic properties for both black holes with graphical methods. }
\end{abstract}



\bigskip

\newpage
\section{Introduction}\label{sec:intro}

Investigation of the gravitational interaction as an intrinsic quantum nature phenomenon has been known as one of the most difficult and exhilarating challenges of physics. For a long time, physicists are looking for a way to construct a flawless quantum theory of gravity. At any energy scale, such a theory has to produce successful predictions without any conceptual debates. Along this line,  many articles that are based on modifying the ordinary Heisenberg uncertainty principle (HUP) on the scale of quantum gravity have been written \cite{1}-\cite{14}. Among the various adopted forms, a particular form that presents a minimal observable length value associated to a string in string theory, namely, generalized uncertainty principle (GUP), is employed in quantum gravity  \cite{1},\cite{2}.  Extended Uncertainty Principle (EUP), which defines a lower bound limit value to the observable momentum quantity in quantum mechanics, is another generalization of the HUP \cite{15}-\cite{24}.   In this context another one is the Doubly Special Relativity (DSR), which takes the Planck energy as a universal constant in addition to the speed of light \cite{25}-\cite{28}.

Recently, it is shown that the GUP also provides an extraordinary understanding of the behavior of particles around black holes \cite{29},\cite{30} as well as their thermodynamic properties \cite{0011}-\cite{16}. To this end, in Ref. \cite{12}, authors considered a special form of the GUP to investigate the thermal properties of a topologically charged black hole (tCBH). In another work, the authors explored the effect of the GUP on the thermodynamic properties of the tCBH in the anti-de Sitter spacetime within the framework of the DSR \cite{13}. Moreover, the authors proposed a new form of the deformation, namely extended GUP (EGUP), to examine the thermodynamics of the Schwarzschild and Reissner–Nordström (RN) black holes in the Snyder-de Sitter model  \cite{14}. Very recently in \cite{15}, the authors used the lower bound limit of {the  EGUP to obtain} the corrected thermodynamic properties of a static black hole.  Then, they compared their findings with the { HUP-formalism  based } results. Similarly, in Ref. \cite{16}, authors considered a new form of the EUP to find out the thermodynamic functions of the Schwarzschild and RN black holes.

{ On the other hand, the main challenge of these suggested deformations is the experimental detection. Over the past decade, articles have appeared in the literature which discuss the observational evidence of quantum gravitational effects \cite{03}-\cite{012}.   For example, in \cite{01} the authors proposed an experimental set-up to probe deformations of canonical commutator of a macroscopic mechanical resonator. In \cite{02}, the authors suggested various speculations on the choice of the priori assumption of the deformation parameters,  and concluded that it might be possible to observe the effects experimentally in the electron tunneling process, determination of Landau energy levels and Hydrogen Lamb shift. In \cite{04} the authors designed an experimental set to observe a change in the dynamics of a mechanical oscillator according to the GUP.  In \cite{05}, at first the authors used the deformed the Schwarzschild metric to obtain the GUP-corrected Hawking temperature, and then, they calculated corrections to the light deflection and perihelion precision. They reported the results in comparison with astronomical measurements and estimated an upper limit value for the deformation parameter.  In a very recent paper \cite{012}, the authors used the previous work as the base and calculated an upper bound value of the GUP-deformation parameter according to the gravitational tests known as the Shapirime delay, gravitational redshift, and geodetic precision. In \cite{06} from a rotating black hole shadow, the author computed an upper bound limit value for the deformation parameter. In \cite{07} and \cite{08} according to the GUP-modified Friedman equations the authors calculated an upper bound value of the deformation parameter by using recent observational data. In \cite{09}, the authors demonstrated that the dimensionless GUP-deformation parameter can be estimated by the gravitational wave event GW150914, which was discovered by the LIGO Scientific and Virgo Collaborations \cite{010},\cite{011}. According to the authors of \cite{can1}, GUP is also can be tested with non-gravitational experiments such as cavity optomechanical setup with ultracold massive molecular oscillators. A summary of the upper bound values of the deformation parameter obtained from various gravitational and non-gravitational experiments are given in detail in \cite{can2}.  }
 
 {Recently, in a very interesting paper \cite{013}, the authors gave a connection between the GUP and the framework of Corpuscular Theory of gravity by examining the modifications in the black hole temperature and evaporation rate.}   In \cite{29}, the authors employed the Hamilton-Jacobi method to examine the quantum gravity effect on the motion of a massive particle that is located near a static black hole. Also, they deformed the Heisenberg algebra, hence the Hamilton-Jacobi equation, to show the effect of the minimum length on the motion of the particle near the cylindrical horizon. Alike,  in \cite{30} the authors revisited the problem by considering different potential energies near a static spherical black hole. They investigated the influence of the minimal length on the motion of massive particles in an unstable equilibrium state and presented a discussion that compares the result of the deformed and undeformed cases. These deformations and their influences can be taken into account to all orders too \cite{30a}-\cite{30c}. It is worth noting that, in some deformed Heisenberg algebra an upper observable limit value to momentum arises. For example,  Pedram {proposed}  a higher-order deformed algebra 
\begin{eqnarray}
	\left[ {X,P} \right] = \frac{{i\hbar }}{{1 - \beta {P^2}}}, \label{s0}
\end{eqnarray} 
{ that is singular at $P=1/\sqrt{\beta}$,} and showed that the semiclassical energy spectrum of the harmonic oscillator is bounded from above \cite{11}. Here, $ \beta $ is the deformation control parameter which is at the scale where the quantum gravitational effects are dominant. This deformed commutation relation produces the following GUP
\begin{eqnarray}
	\Delta X\Delta P \geqslant \frac{\hbar }{2}\frac{1}{{1 - \beta \left[ {\Delta {P^2} + {{\left\langle \hat P \right\rangle }^2}} \right]}}.
	\label{s1}
\end{eqnarray}
The expectation value of the momentum 
{ is a constant which can take any value regarding the wave function. In order to avoid a shift, which is not physically meaningful, we} set $ \left\langle {\left| {\hat p} \right|} \right\rangle =0${, by considering mirror symmetric states,} and express Eq. \eqref{s1} as follows:
\begin{eqnarray}
	\Delta X\Delta P \geqslant \frac{\hbar }{2}\frac{1}{{1 - \beta \Delta {P^2}}} .
	\label{s2}
\end{eqnarray}
We would like to emphasize that this GUP form also predicts a minimal length uncertainty  \cite{30e}, and it is consistent with the various proposals of quantum gravity such as string theory, loop quantum gravity, DSR and noncommutative space-time \cite{31}-\cite{40}.

In this manuscript, we intend to comprehend the thermal properties of the Schwarzschild and RN black holes in the presence of minimal length and maximum observable momentum bounds. Therefore, we take the deformed higher-order GUP relation which is given in Eq. \eqref{s0} into account and investigate the mass-temperature, entropy, and heat capacity functions of both black holes. We construct the manuscript as follows: In the next two sections, we briefly introduce some features of black hole thermodynamics.  After that,  we obtain the corrected temperature-mass, entropy, and heat capacity relations of the Schwarzschild black hole. Alike, in the next section, we examine the RN charged black hole's corrected thermodynamic properties. Finally, we conclude this paper with a short conclusion in the last section.

\section{Black hole thermodynamics}
\label{sec:Black hole thermodynamics}
\hspace{0.5cm}
According to Bekenstein, entropy function of a black hole has to be proportional with the horizon area, $A$,  \cite{bek73}. In the semi-classical case, one can express entropy and temperature of a black hole as {
\begin{eqnarray}
	{S_0} = \frac{A}{{2\hbar }},\,\,\,\,\,\,\,\,\,\,\,\,{T_0} = \frac{{\kappa \hbar}}{{4\pi }},\label{e1}
\end{eqnarray}}
where $\kappa$ represents the surface gravity \cite{13}. If a black hole absorbs a particle, the smallest possible change in its area{, $\Delta A$,} can be taken proportional to the size and the mass of the particle as
\begin{eqnarray}
	{\Delta A} \sim Xm, \label{s2a}
\end{eqnarray}
that leads to ${\left( {\Delta S} \right)_{\min }} = \rm \ln 2$  \cite{xiang}. One can associate the uncertainty in position with the size of the particle (${X \sim \Delta X}$). Besides, we know that the uncertainty in momentum should not exceed the mass value (${\Delta P \leqslant m}$). According to these facts, one can rewrite horizon area in the following form:
\begin{eqnarray}
	{ A  \geqslant \Delta X\Delta P.}\label{s6}
\end{eqnarray}
When the black hole absorbs the particle, $\Delta X$ should not be greater than a particular scale which minimizes the value of $\Delta A$. This particular size should be related to the properties of the black hole, whether $\Delta A_{\rm min}$ is being expected to represent an intrinsic property of the black hole's horizon. 

\section{Thermodynamics of the Schwarzschild Black Hole}
\label{sec:Thermodynamics of the Schwarzschild Black Hole}
At first, we consider a static spherically symmetric Schwarzschild black hole whose size is defined by the horizon radius given below 
\begin{eqnarray}
	\frac{\hbar }{{2\Delta P}}\left( {\frac{1}{{1 - \beta \,\Delta {P^2}}}} \right) \leqslant \Delta X \leqslant 2{r_0}. \label{s4}
\end{eqnarray}
By solving Eq. \eqref{s4}, we find the momentum uncertainty as
\begin{equation}
	\begin{split}
		&\Delta P \geqslant \frac{{2{r_0}}}{{{3^{1/3}}{{\left( { - 9\hbar {r_0}^2{\beta ^2} + \sqrt 3 \sqrt { - 64{r_0}^6{\beta ^3} + 27{\hbar ^2}{r_0}^4{\beta ^4}} } \right)}^{1/3}}}}\\
		& + \frac{{{{\left( { - 9\hbar {r_0}^2{\beta ^2} + \sqrt 3 \sqrt { - 64{r_0}^6{\beta ^3} + 27{\hbar ^2}{r_0}^4{\beta ^4}} } \right)}^{1/3}}}}{{2 \times {3^{2/3}}{r_0}\beta }}, \label{s5}
	\end{split}
\end{equation}
Then, we substitute Eqs. \eqref{s4} and \eqref{s5} in Eq. \eqref{s6}, and express the increase of the horizon area with the multiplication of the calibration factor and the corrected Planck constant as
\begin{eqnarray}
	{\Delta} A \geqslant \eta \hbar', \label{s61}
\end{eqnarray}
where 
\begin{eqnarray}
	\hbar' = \frac{\hbar }{{2}}\left[ {1 - \frac{{{{\left( {4 \times {3^{2/3}}{r_0}^2\beta  + {3^{1/3}}{{\left( { - 9\hbar r_{0}^2{\beta ^2} + \sqrt 3 \sqrt {{r_0}^4{\beta ^3}\left( { - 64{r_0}^2 + 27{\hbar ^2}\beta } \right)} } \right)}^{2/3}}} \right)}^2}}}{{36{r_0}^2\beta {{\left( { - 9\hbar {r_0}^2{\beta ^2} + \sqrt 3 \sqrt {{r_0}^4{\beta ^3}\left( { - 64{r_0}^2 + 27{\hbar ^2}\beta } \right)} } \right)}^{2/3}}}}} \right]^{-1}.
	\label{s8}
\end{eqnarray}
It is worth noting that in the absence of the deformation parameter, namely for $ \beta = 0$,  the GUP reduces to the HUP and $ \hbar'$ becomes equal to $\dfrac{\hbar}{2}$. According to this modification, we immediately express the corrected temperature function in the semiclassical framework as
\begin{eqnarray}
	\label{g2}
	T' &=& \frac{{\kappa \hbar'}}{{2\pi }}
	\nonumber\\
	&=& \frac{{\hbar \kappa }}{{4\pi }}\left[ {1 - \frac{{{{\left( {4 \times {3^{2/3}}{r_0}^2\beta  + {3^{1/3}}{{\left( { - 9\hbar {r_0}^2{\beta ^2} + \sqrt 3 \sqrt {{r_0}^4{\beta ^3}\left( { - 64{r_0}^2 + 27{\hbar ^2}\beta } \right)} } \right)}^{2/3}}} \right)}^2}}}{{36{r_0}^2\beta {{\left( { - 9\hbar {r_0}^2{\beta ^2} + \sqrt 3 \sqrt {{r_0}^4{\beta ^3}\left( { - 64{r_0}^2 + 27{\hbar ^2}\beta } \right)} } \right)}^{2/3}}}}} \right]^{-1}.\label{s9}
\end{eqnarray}
Note that for $ \beta=0 $, Eq. (\ref{s9}) reduces to $T' =  \frac{{\kappa \hbar}}{{4\pi }}$ which is the same with Eq. (\ref{e1}), and moreover with findings in the absence of correction terms of the considered GUP of \cite{14}.

Then, we derive the corrected heat capacity function of the GUP black hole from the usual well-known analytical formalism:
\begin{eqnarray}
	C' = {c^2}\frac{{dM}}{{dT'}} = {\left( {\frac{1}{{{c^2}}}\frac{{dT'}}{{dM}}} \right)^{ - 1}}.\label{s16}
\end{eqnarray}
Here, we take the corrected temperature function up to the first order of the GUP parameter terms 
\begin{eqnarray}
	T'(\beta) = \frac{{\hbar \kappa }}{{4\pi }} + \frac{{{\hbar ^3}\kappa \beta }}{{64\pi {r_0}^2}} + O{\left[ \beta  \right]^2}. \label{s17}
\end{eqnarray}
Next, we employ the radius of the Schwarzschild black hole radius  ${r_0} = \frac{{2GM}}{{{c^2}}}$ in Eq. \eqref{s17}. Following the simple algebra of Eq. \eqref{s16}, we find the corrected heat capacity function in the form of
\begin{align}
	C'(\beta) =  - \frac{128 \pi G^2 M^3 }{c^2 \beta \kappa \hbar ^3}, 
\end{align}
Instead, if we take the corrected temperature function up to the second order of the GUP parameter
\begin{eqnarray}
	T'(\beta^{2}) = \frac{{\hbar \kappa }}{{4\pi }} + \frac{{{\hbar ^3}\kappa \beta }}{{64\pi {r_0}^2}} + \frac{{3{\hbar ^5}\kappa {\beta ^2}}}{{1024\pi {r_0}^4}} + O{\left[ \beta  \right]^3}, \label{s19}
\end{eqnarray}
we arrive at 
\begin{align}
	C' (\beta^{2})=  - \frac{{4096 \pi {G^4}{M^5} }}{{32{c^2}{G^2}{M^2}\beta \kappa {\hbar ^3} + 3{c^6}{\beta ^2}\kappa {\hbar ^5}}}. 
\end{align} 
We examine entropy function as the last thermal property. For this, we take the following analytical method into account to express the modified entropy of the black hole.
\begin{eqnarray}
	S' = \int {\frac{{dS}}{{dA}}} dA \simeq \int {\frac{{{{\left( {\Delta S} \right)}_{\min }}}}{{{{\left( {\Delta A} \right)}_{\min }}}}} dA \simeq \int {\frac{{dA}}{{4\hbar '}}} ,\label{s10}
\end{eqnarray}
According to Eq. \eqref{s61} and Ref. \cite{xiang}, we find  
\begin{eqnarray}
	\frac{{dA}}{{dS}} = \frac{{{{\left( {\Delta A} \right)}_{\min }}}}{{{{\left( {\Delta S} \right)}_{\min }}}} = 4\hbar',\label{s11}
\end{eqnarray}
where we take the calibration factor as $\eta  = 4\ln 2$. With the heuristic approach, 
\begin{eqnarray}
	\hbar '(\beta) = \frac{\hbar }{2} + \frac{{{\hbar ^3}\beta }}{{32{r_0}^2}} + O{\left[ \beta  \right]^2} ,\label{s12}
\end{eqnarray}
we calculate the modified entropy of the black hole up to the first order of the GUP parameter as
\begin{eqnarray}
	S'(\beta)= \frac{{2\pi {r_0}^2}}{\hbar } - \frac{{\hbar \pi \beta }}{8} Ln\left[ {16{r_0}^2 + {\hbar ^2}\beta } \right]. \label{s13}
\end{eqnarray}
We observe that the GUP modification produces a logarithmic term in entropy function similar to many other quantum corrected entropy functions.

Instead if we take the modification up to the second order of the GUP parameter, we get
\begin{eqnarray}
	\hbar '(\beta^{2}) = \frac{\hbar }{2} + \frac{{{\hbar ^3}\beta }}{{32{r_0}^2}} + \frac{{3{\hbar ^5}{\beta ^2}}}{{512\left( {{r_0}^4} \right)}} + O{\left[ \beta  \right]^3} ,\label{s14}
\end{eqnarray}
and hence, 
\small{
\begin{eqnarray}
	S'(\beta^{2}) = \frac{{2\pi {r_0}^2}}{\hbar } - \frac{{\hbar \pi \beta }}{{176}}\left( {10\sqrt {11} Arctan\left[ {\frac{{32{r_0}^2 + {\hbar ^2}\beta }}{{\sqrt {11} \,{\hbar ^2}\beta }}} \right] + 11Ln\left[ {256{r_0}^4 + 16{\hbar ^2}{r_0}^2\beta  + 3{\hbar ^4}{\beta ^2}} \right]} \right).\,\,\label{s15}
\end{eqnarray}}
\normalsize 
It is worth noting that in the limit of $ \beta\rightarrow0 $, the modified entropy functions, namely Eqs. \eqref{s13} and \eqref{s15}, reduce to the ordinary Schwarzschild entropy function, Eq. \eqref{e1}, which is obtained in the semiclassical framework.

\section{Thermodynamics of RN Black Hole}\label{sec:sec4}
The RN black hole differs from the  Schwarzschild black hole in terms of being charged and having two possible horizon radius. Therefore, we think it would be interesting to examine the GUP corrected thermal properties of the RN black hole. We start by defining the metric of the RN black hole \cite{13}
\begin{eqnarray}
	\label{t1}
	ds^{2}=f(r)dt^{2}-\dfrac{dr^{2}}{f(r)}-r^{2}(d\theta^{2}+sin^{2}\theta d\varphi^{2}), \label{n1}
\end{eqnarray}
where
\begin{eqnarray}
	\label{t2}
	f(r)=1-\frac{2M}{r}+\frac{Q^{2}}{r^{2}}. \label{n2}
\end{eqnarray}		
Here, $M$ and $Q$ denote the mass and charge, respectively. Two possible horizon radius of the RN black hole are obtained from the roots of Eq. \eqref{n2} as $r_{\pm}= M \pm \sqrt {{M^2} - {Q^2}}$. Note that in the zero-charge limit, these radii reduce to the Schwarzschild black hole radius. Since the radius has to be a real number, a constraint arises on the mass and charge, such as $0<Q<M$, and thus, the event horizon splits into an inner and an outer horizon. In that case, the outer horizon radius of the RN black hole is written as
${r_{RN}} = \frac{{G{r}}}{{{c^2}}}$ \cite{14}, where $r$ is
\begin{eqnarray}
	\label{f8}
	{r} = M + \sqrt {{M^2} - {Q^2}}.\label{n3}
\end{eqnarray}
During a particle absorption in the RN black hole, the particle's position uncertainty is assumed to be proportional to its outer horizon radius \cite{14}
\begin{eqnarray}
	\label{f9}
	\Delta X = \gamma {r_{RN}},\label{n4}
\end{eqnarray}
This proportionality is expressed with a scale factor, $\gamma $. On the other hand momentum uncertainty can be associated to the temperature, $T$, with
\begin{eqnarray}
	\label{j1}
	T = \frac{{c\Delta P}}{\kappa_B }. \label{n6}
\end{eqnarray}
Here, the Boltzmann constant is represented by $\kappa_B$ \cite{14}. Then, we use Eqs. \eqref{n4} and \eqref{n6} with the relation, ${\left( {{m_p}c} \right)^2} = \frac{{\hbar {c^3}}}{G}$,  in between the Planck mass, $m_P$, and the Netwon's universal gravitational constant, $G$, to rewrite Eq. \eqref{s2}. We get
\begin{eqnarray}
	\label{f10}
	{r} \geqslant \frac{{{{({m_p}c)}^2}}}{{2\gamma \kappa_B T}}\bigg[\frac{1}{{1 - \beta \Delta {P^2}}}\bigg].\label{n7}
\end{eqnarray}
In order to determine the scale factor, we first take the temperature of the RN black hole to the HUP limit. We find $T = \frac{{{{({m_p}c)}^2}}}{{2\gamma \kappa_B {r}}}$. Then, we match this scale factor dependent temperature with the semiclassical Hawking temperature,  ${T_H} = \frac{{{{({m_p}c)}^2}\left( {M{r} - {Q^2}} \right)}}{{2\pi \kappa_B r^3}}$. So that, we find out the scale factor as:
\begin{eqnarray}
	\gamma  = \frac{{\pi r^2}}{{M{r} - }{Q^{2}}}.\label{n8}
\end{eqnarray}
Next, we derive the mass-charge-temperature equation of the RN black hole by substituting Eqs. \eqref{f8} and \eqref{n8} into Eq. \eqref{n7}. We get
\begin{eqnarray}
	M + \sqrt {{M^2} - {Q^2}}  - \frac{(m_Pc^2)^2}{2\pi\kappa_B} \frac{\sqrt {{M^2} - {Q^2}}}{M+\sqrt {{M^2} - {Q^2}}}\frac{1}{\left( {T^2}\beta {\kappa_B^2}- {c^2}\right)T}=0. \label{n9}
\end{eqnarray}
This equation has three roots, but two of them are complex-valued expressions, so they are unphysical. The only physical root in its closed form (according to Eq. \eqref{n7})  is as follows:
\small
\begin{eqnarray}
	{T_{RN}} &=&  \frac{{{2^{2/3}}{c^2}\gamma r\kappa_B }}{{{3^{1/3}}{{\left( {-9{c^4}{\gamma ^2}{m_p}^2{r^2}{\beta ^2}{\kappa_B^6} + \sqrt 3 \sqrt {  27{c^8}{\gamma ^4}{m_p}^4{r^4}{\beta ^4}{\kappa_B^{12}}- 16{c^6}{\gamma ^6}{r^6}{\beta ^3}{\kappa_B^{12}} } } \right)}^{1/3}}}} \nonumber\\
	&+& \frac{{{{\left( {-9{c^4}{\gamma ^2}{m_p}^2{r^2}{\beta ^2}{\kappa_B^6} + \sqrt 3 \sqrt {  27{c^8}{\gamma ^4}{m_p}^4{r^4}{\beta ^4}{\kappa_B^{12}}- 16{c^6}{\gamma ^6}{r^6}{\beta ^3}{\kappa_B^{12}}} } \right)}^{1/3}}}}{{{6^{2/3}}\gamma r\beta {\kappa_B^3}}}.
	\label{n10}
\end{eqnarray}
\normalsize
Next, we derive heat capacity function with the help of the mass-charge-temperature equation. First, we use the Taylor expansion of Eq. \eqref{n10} up to the first  order of the deformation parameter. We find
\begin{eqnarray}
	{T_{RN}}(\beta) &=& \frac{{{c^2}{m_p}^2}}{{2\gamma r\kappa_B }} + \frac{{{c^4}{m_p}^6\beta }}{{8{\gamma ^3}{r^3}\kappa_B }} + O{\left[ \beta  \right]^2},   \nonumber\\
	&=& \frac{{{c^2}{m_p}^2\sqrt {{M^2} - {Q^2}} }}{{2\pi {{\left( {M + \sqrt {{M^2} - {Q^2}} } \right)}^2}\kappa_B }} + \frac{{{c^4}{m_p}^6{{\left( {{M^2} - {Q^2}} \right)}^{3/2}}\beta }}{{8{\pi ^3}{{\left( {M + \sqrt {{M^2} - {Q^2}} } \right)}^6}\kappa_B }} + O{\left[ \beta  \right]^2}. \label{n11}
\end{eqnarray}
According to Eq. \eqref{s16}, we derive the first-order corrected heat capacity function  in the form of
\small
\begin{eqnarray}
	{C_{RN}}(\beta) &=&  - \left( {8{\pi ^3}\sqrt {{M^2} - {Q^2}} {{\left( {M + \sqrt {{M^2} - {Q^2}} } \right)}^6}\kappa_B } \right)\bigg[{m_p}^2\bigg(32{M^5}{\pi ^2}  
	+ 32{M^4}{\pi ^2}\sqrt {{M^2} - {Q^2}} 
	\nonumber\\
	& +& 2{Q^2}\sqrt {{M^2} - {Q^2}} \left( {4{\pi ^2}{Q^2} - 3{c^2}{m_p}^4\beta } \right)  
	+ 6{M^2}\sqrt {{M^2} - {Q^2}} \left( { - 8{\pi ^2}{Q^2} + {c^2}{m_p}^4\beta } \right) 
	\nonumber\\
	& -& {M^3}\left( {64{\pi ^2}{Q^2} + 3{c^2}{m_p}^4\beta } \right)  
	+ M\left( {28{\pi ^2}{Q^4} + 3{c^2}{m_p}^4{Q^2}\beta } \right)\bigg){\bigg]^{ - 1}}. \label{n12}
\end{eqnarray}
\normalsize
Similarly, in order to obtain the second-order corrected heat function we first expand Eq. \eqref{n11} up to second order of $\beta$
\begin{eqnarray}
	{T_{RN}}(\beta^{2}) &=& \frac{{{c^2}{m_p}^2}}{{2\gamma r\kappa_B }} + \frac{{{c^4}{m_p}^6\beta }}{{8{\gamma ^3}{r^3}\kappa_B }} + \frac{{3{c^6}{m_p}^{10}{\beta ^2}}}{{32{\gamma ^5}{r^5}\kappa_B }} + O{\left[ \beta  \right]^3}  
	\nonumber\\
	&=& \frac{{{c^2}{m_p}^2\sqrt {{M^2} - {Q^2}} }}{{2\pi {{\left( {M + \sqrt {{M^2} - {Q^2}} } \right)}^2}\kappa_B }} + \frac{{{c^4}{m_p}^6{{\left( {{M^2} - {Q^2}} \right)}^{3/2}}\beta }}{{8{\pi ^3}{{\left( {M + \sqrt {{M^2} - {Q^2}} } \right)}^6}\kappa_B }} 
	\nonumber\\
	&+& \frac{{3{c^6}{m_p}^{10}{{\left( {{M^2} - {Q^2}} \right)}^{5/2}}{\beta ^2}}}{{32{\pi ^5}{{\left( {M + \sqrt {{M^2} - {Q^2}} } \right)}^{10}}\kappa_B }} + O{\left[ \beta  \right]^3} , \label{n13}
\end{eqnarray}
and then, accordingly, we find
\small
\begin{eqnarray}
	{C_{RN}}(\beta^{2}) &=& \left( {32{\pi ^5}\sqrt {{M^2} - {Q^2}} {{\left( {M + \sqrt {{M^2} - {Q^2}} } \right)}^{10}}\kappa_B } \right)
	\bigg[c\bigg(16M{\pi ^4}{\left( {M + \sqrt {{M^2} - {Q^2}} } \right)^8} 
	\nonumber\\ 
	&-& 32{\pi ^4}\sqrt {{M^2} - {Q^2}} {\left( {M + \sqrt {{M^2} - {Q^2}} } \right)^8} + 12{c^2}Mm_{p}^4{\pi ^2}\left( {{M^2} - {Q^2}} \right){\left( {M + \sqrt {{M^2} - {Q^2}} } \right)^4}\beta 
	\nonumber\\ 
	&-& 24{c^2}m_{p}^4{\pi ^2}{\left( {{M^2} - {Q^2}} \right)^{3/2}}{\left( {M + \sqrt {{M^2} - {Q^2}} } \right)^4}\beta  + 15{c^4}Mm_{p}^8{\left( {{M^2} - {Q^2}} \right)^2}{\beta ^2}
	\nonumber\\   
	&-& 30{c^4}m_{p}^8{\left( {{M^2} - {Q^2}} \right)^{5/2}}{\beta ^2}\bigg){\bigg]^{ - 1}}.\label{n14}
\end{eqnarray}	 
\normalsize
Next, we intend to derive the GUP-corrected entropy function. We employ the definition given in Eq. \eqref{s10}. After following the same methodolgy, we find $\hbar'$ for the RN black hole as
\small
\begin{eqnarray}
	\hbar '&=&\frac{\hbar }{2}\left[ {1 - \frac{{{{\left( {{{43}^{2/3}}{r_{RN}}^2\beta  + {3^{1/3}}{{\left( { - 9\hbar {r_{RN}}^2{\beta ^2} + \sqrt 3 \sqrt {{r_{RN}}^4{\beta ^3}\left( { - 64{r_{RN}}^2 + 27{\hbar ^2}\beta } \right)} } \right)}^{2/3}}} \right)}^2}}}{{36{r_{RN}}^2\beta {{\left( { - 9\hbar {r_{RN}}^2{\beta ^2} + \sqrt 3 \sqrt {{r_{RN}}^4{\beta ^3}\left( { - 64{r_{RN}}^2 + 27{\hbar ^2}\beta } \right)} } \right)}^{2/3}}}}} \right]^{-1}
	\nonumber\\
	&=& \frac{\hbar }{2}\Bigg[1 - \beta \Bigg(\frac{{2Gr}}{{{3^{1/3}}{{\left( { - 9{c^2}{G^2}\hbar {r^2}{\beta ^2} + \sqrt 3 \sqrt {{G^4}{r^4}{\beta ^3}\left( { - 64{G^2}{r^2} + 27{c^4}{\hbar ^2}\beta } \right)} } \right)}^{1/3}}}} \nonumber\\
	& +& \frac{{{{\left( { - 9{c^2}{G^2}\hbar {r^2}{\beta ^2} + \sqrt 3 \sqrt {{G^4}{r^4}{\beta ^3}\left( { - 64{G^2}{r^2} + 27{c^4}{\hbar ^2}\beta } \right)} } \right)}^{1/3}}}}{{{{23}^{2/3}}Gr\beta }}{\Bigg)^2}{\Bigg]^{- 1}},\label{n16}
\end{eqnarray}
\normalsize
We observe that in the absence of the deformation parameter, $\hbar$ in the above equation  reduces to $\dfrac{\hbar}{2}$. For the first order corrected-entropy function, we consider the Taylor expansion of $ \hbar ' $ 
\begin{eqnarray}
	\hbar '(\beta) = \frac{\hbar }{2} + \frac{{{\hbar ^3}\beta }}{{32{r_{RN}}^2}} + O{\left[ \beta  \right]^2} ,\label{n17}
\end{eqnarray}
and express the entropy as follows:
\begin{eqnarray}
	S_{RN}(\beta)= \frac{{2{G^2}\pi {{\left( {M + \sqrt {{M^2} - {Q^2}} } \right)}^2}}}{{{c^4}\hbar }} - \frac{1}{8}\pi \beta \hbar Ln\left[ {\frac{{16{G^2}{{\left( {M + \sqrt {{M^2} - {Q^2}} } \right)}^2}}}{{{c^4}}} + \beta {\hbar ^2}} \right],\label{n18}
\end{eqnarray}
To derive the second order corrected entropy, we first use the following expansion, 
\begin{eqnarray}
	\hbar '(\beta^{2}) = \frac{\hbar }{2} + \frac{{{\hbar ^3}\beta }}{{32{r_{RN}}^2}} + \frac{{3{\hbar ^5}{\beta ^2}}}{{512{r_{RN}}^4}} + O{\left[ \beta  \right]^3},\label{n19}
\end{eqnarray}
then, we obtain
\small
\begin{eqnarray}
	S_{RN}(\beta^{2})& =&
	\frac{{2\pi r_{RN}^2}}{\hbar } - \frac{\pi \beta \hbar}{{176}} \bigg(10\sqrt {11} Arctan\left[ {\frac{{{{32{G^2}{{\left( {M + \sqrt {{M^2} - {Q^2}} } \right)}^2}}} + \beta {\hbar ^2}c^4}}{{\sqrt {11} \beta {\hbar ^2}c^4}}} \right]  \nonumber\\ 
	&+& 11Ln\left( {\frac{{256{G^4}{{\left( {M + \sqrt {{M^2} - {Q^2}} } \right)}^4}}}{{{c^8}}} + \frac{{16{G^2}{{\left( {M + \sqrt {{M^2} - {Q^2}} } \right)}^2}\beta {\hbar ^2}}}{{{c^4}}} + 3{\beta ^2}{\hbar ^4}}\right)\Bigg] \,\,\,\,\,\,\,\,\,\, .\label{n20}
\end{eqnarray}
\normalsize
By considering $ \beta\rightarrow0 $ limit, we observe that the GUP corrected entropy up to the first order, Eq. \eqref{n18}, and up to the second order, Eq. \eqref{n20}, become identical to the black hole entropy, given in Eq. \eqref{e1}, which is produced in the semiclassical framework. Moreover, as in the case of the Schwarzschild black hole, {the} GUP correction leads to additional logarithmic terms to the entropy function.   

Before concluding the manuscript, we would like to plot the thermal functions in order to provide a better understanding. During this analysis, we take $G=m_P=c=\hbar=\kappa_B=1$. 

\begin{figure}[htb!]
	\includegraphics[height=8cm,width=13cm]{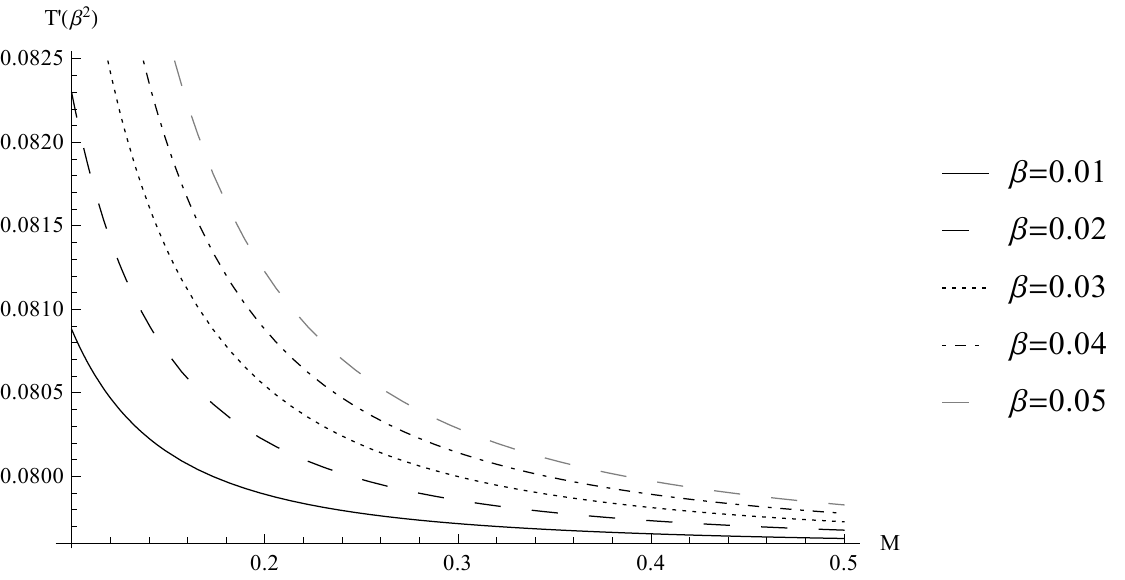}
	\caption{The second-order GUP-corrected temperature versus the mass of the Schwarzschild black hole for various $\beta$. }
	\label{fig:fig1}
\end{figure}

\begin{figure}[htb!]
	\includegraphics[height=8cm,width=13cm]{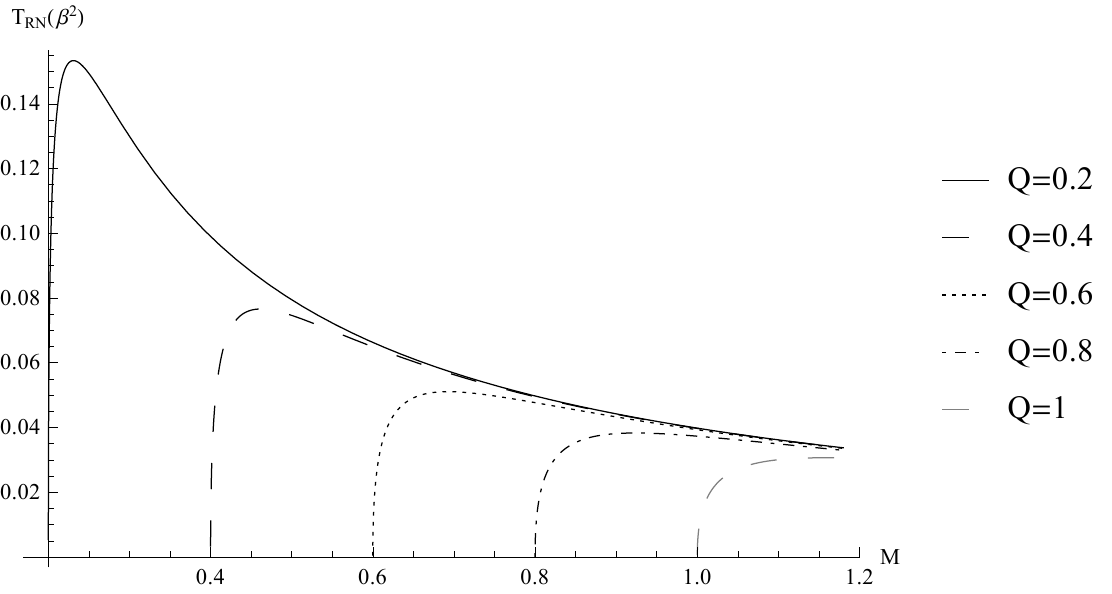}
	\caption{The second-order GUP-corrected temperature versus the mass of variously charged RN black holes for $\beta=0.05$. }
	\label{fig:fig2}
\end{figure}

We start with the illustration of the mass-temperature functions.  In Fig. (\ref{fig:fig1}), We depict the Schwarzschild black hole's case. When the mass increases, we observe a decrease in temperature function, too. {However, this decrease relatively becomes greater for higher values of the deformation parameter.} In Fig. (\ref{fig:fig2}) we demonstrate the {charged} RN black hole's case {by taking deformation value as $\beta=0.05$.} We observe that at certain mass values, the temperature rapidly increases, and then, relaxes slowly. These certain mass values depend on the charge and for smaller charge values the temperature function characteristic tends to mimic the Schwarzschild black hole's one.  {In both figures, we observe that the black hole's temperature is bounded from above as a natural consequence of the employed GUP formalism.}

\begin{figure}
	\includegraphics[height=8cm,width=13cm]{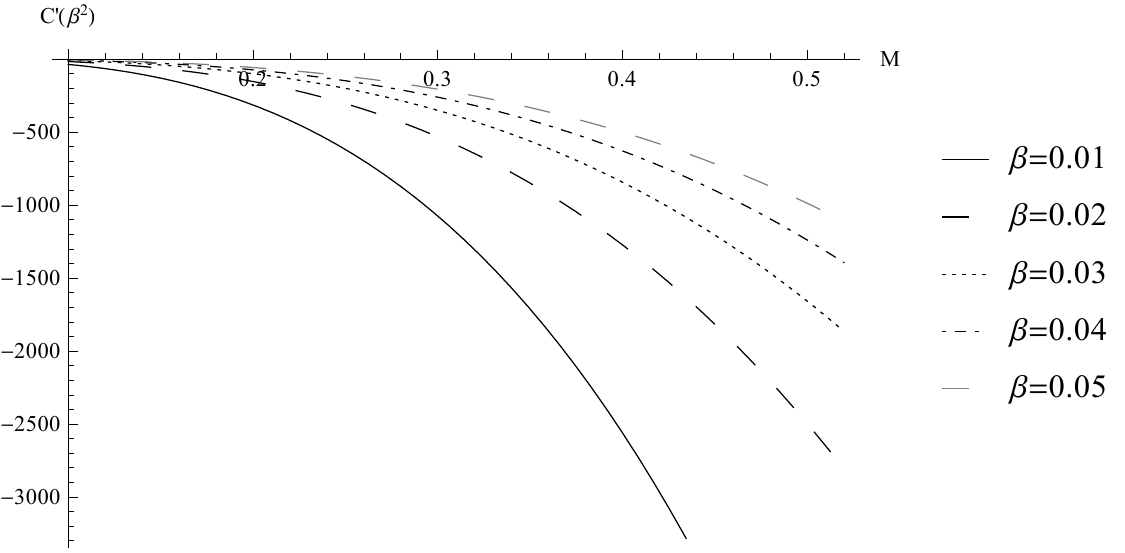}
	\caption{The second-order GUP-corrected specific heat versus the mass of the Schwarzschild black hole for various $\beta$.}
	\label{fig:fig3}
\end{figure}

\begin{figure}[htb!]
	\includegraphics[height=8cm,width=13cm]{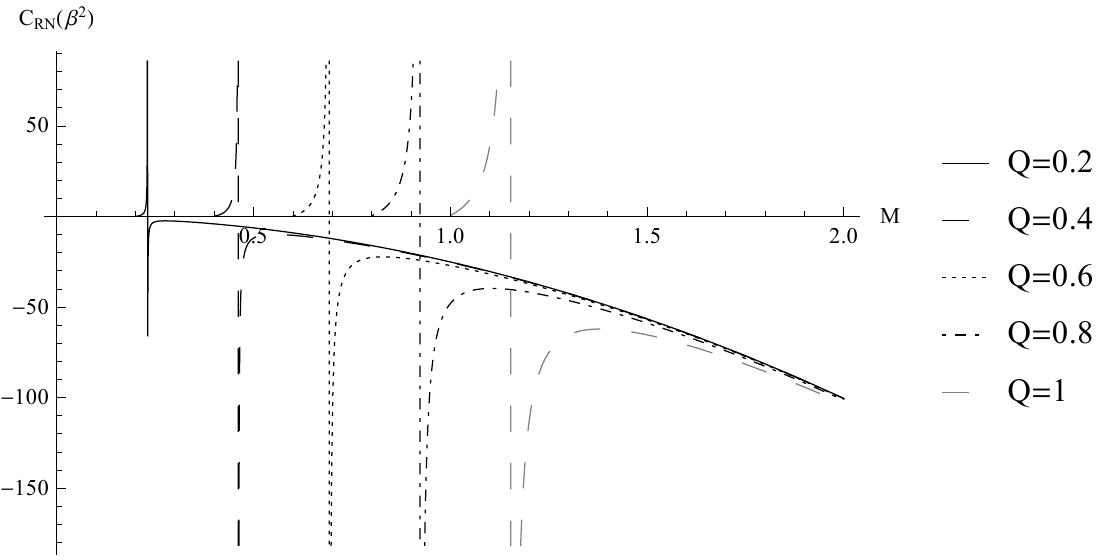}
	\caption{The second-order GUP-corrected specific heat versus the mass of variously charged RN black holes for $\beta=0.05$.}
	\label{fig:fig4}
\end{figure}

We continue with the illustration of the heat capacity functions. In Figs. (\ref{fig:fig3}) and (\ref{fig:fig4}) we depict the Schwarzschild and the RN black hole heat capacity functions via the mass, respectively. We find a negative heat capacity in the Schwarzschild as expected. However, we notice that the deformation parameter changes the speed of decrements. In the RN black hole case, we use a fixed deformation value as we have done above.  We observe that the mass increase causes a rapid increase in the heat capacity function at first, then after a critical mass value, it leads to a relatively slow decrease. Furthermore, when the mass becomes much bigger than the charge, the specific heat functions become identical to each other. 

\begin{figure}[htb!]
	\includegraphics[height=8cm,width=13cm]{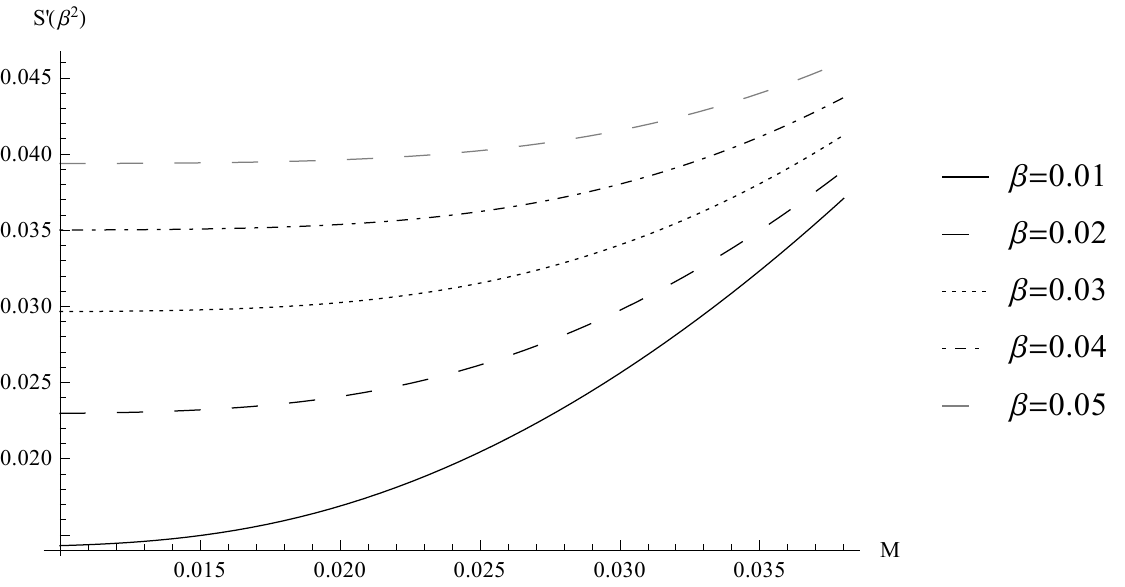}
	\caption{The second-order GUP-corrected entropy versus the mass of the Schwarzschild black hole for various $\beta$.}			
	\label{fig:fig5}
\end{figure}

\begin{figure}[htb!]
	\includegraphics[height=8cm,width=13cm]{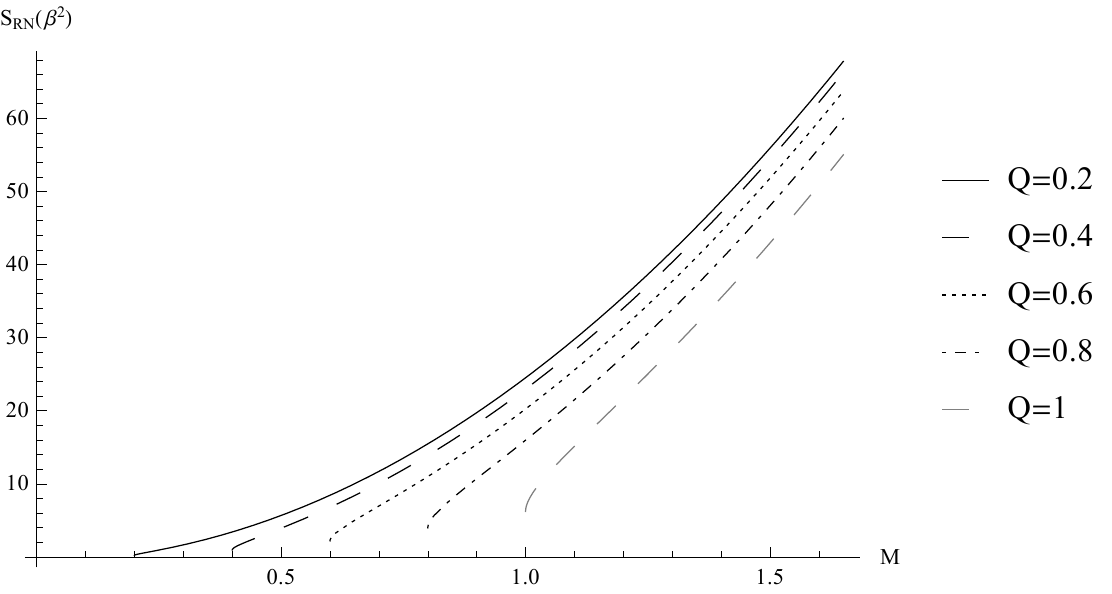}
	\caption{The second-order GUP-corrected specific heat versus the mass of  variously charged RN black holes for $\beta=0.05$.}
	\label{fig:fig6}
\end{figure}

We end up with the demonstration of the entropy functions. In Fig. (\ref{fig:fig5}), we depict the second-order corrected-entropy function of the Schwarzschild black hole via five different deformation parameters. We observe an increase in entropy due to the increase in mass. However, we say that this increase is sharper with a small deformation parameter value. In Fig. (\ref{fig:fig6}), we present the case in the RN black hole. For the fixed deformation parameter, we observe a relatively smaller entropy value for higher charge values.




\section{Conclusion\label{Conc}}
\label{sec:Conclusions}
In this manuscript, we examined the thermodynamics properties of the Schwarzschild and Reissner-Nordst\"orm (RN) black holes by considering a higher-order generalized uncertainty principle (GUP) instead of the ordinary Heisenberg uncertainty principle (HUP).  {At first,} we derived the corrected temperature-mass relations for the Schwarzschild black hole. {Then}, we {derived} the corrected mass-temperature–charge relation for the RN black hole. {We found an upper bound for the black hole's temperature which is consistent with the used deformed algebra.} In the semiclassical framework,  we derived the GUP-modified entropy in terms of mass for the Schwarzschild black hole and in terms of mass and charge for the RN black hole, respectively. Also, we calculated the GUP-modified heat capacity for the Schwarzschild and charged black holes. We studied all these thermal properties up to the first and second-order of the deformation parameter.  {Finally}, we compared all these results with each other { by graphical demonstration. We confirmed our findings by considering them in the HUP limits which already exist in the literature. }

\section*{Acknowledgements}
{The authors thank the anonymous reviewer for his/her helpful and constructive comments.} This work is supported by the Internal Project, [2021/2212], of Excellent Research of the Faculty of Science of University Hradec Kr\'alov\'e.





\newpage

\end{document}